\documentclass[final]{eps}

\usepackage{graphicx}
\usepackage{times}

\volumenumber{58}
\publishyear{2006}
\frompage{e41}
\topage{e44 }

\shorttitle{
T. Umeda and R. Yamazaki: 
A shock-rest-frame model
}

\title{
Full particle simulation of a perpendicular collisionless shock: \\
A shock-rest-frame model
}

\author{
Takayuki Umeda$^1$
and 
Ryo Yamazaki$^2$
}

\affiliation{
$^1$Solar-Terrestrial Environment Laboratory, Nagoya University, 
Nagoya, Aichi 464-8601, JAPAN \\
$^2$Department of Physics, Hiroshima University, 
Higashi-Hiroshima, Hiroshima 739-8526, JAPAN
}

\received{May 25, 2006}
\revised{July 20, 2006}
\accepted{July 25, 2006}

\abstract{
The full kinetic dynamics of a 
perpendicular collisionless shock is studied 
by means of a one-dimensional electromagnetic full particle simulation. 
The present simulation domain is taken in the shock rest frame
in contrast to the previous full particle simulations of shocks. 
Preliminary results show that the downstream state 
falls into a unique cyclic reformation state 
for a given set of upstream parameters 
through the self-consistent kinetic processes. 
}

\keywords{
Collisionless shock; particle-in-cell simulation; 
shock rest frame
}

\begin{document}
\label{firstpage}
\maketitle
\copyrighttext{}
%\clearpage

%%%%%%%%%%%%%%%%%%%%%%%%%%%%%%%%%%%%%%%%%%%%%%%%%%%%%%%%%%%%%%%%%%%%%

\section{Introduction}

Collisionless shocks are universal processes in space and 
are observed in laboratory, astrophysical, and space plasmas, 
including astrophysical jets, an interstellar medium, 
the heliosphere, and the planetary magnetosphere. 
The classic picture of collisionless shocks is that they
arise in a
collisionless plasma streaming with a supersonic velocity 
against another collisionless plasma. 
The generation of plasma waves and turbulences, plasma heating and 
acceleration, and electromagnetic radiation processes 
at collisionless shocks are of great interest 
in astrophysics, solar physics, planetary physics, geophysics, 
and plasma physics. 
Detailed scientific issues in collisionless shock physics 
are described in Lembege \textit{et al.} (2004). 
Kinetic simulations of collisionless shocks are essential
approaches to solve these scientific issues.

% Methods for exciting shocks

There are several different methods for exciting 
collisionless shocks in numerical simulations of plasmas. 
The magnetic piston method 
(Lembege and Dawson, 1987a; Lembege and Savoini, 1992) 
is widely used in full particle simulations, 
in which a plasma is accelerated 
by an external current pulse applied at one side 
of the simulation domain. 
The simulation domain is taken in the upstream rest frame. 
The plasma is pushed by the ``magnetic piston'' 
into the background plasma, 
and the external pulse develops into a shock wave. 
Another method widely used is the injection method 
(Quest, 1985; Shimada and Hoshino, 2000; 
 Schmitz \textit{et al.}, 2002a, b; Nishimura \textit{et al.}, 2003; 
 Scholer \textit{et al.}, 2003; Lee \textit{et al.}, 2004), 
in which a plasma is injected from one side 
of the simulation domain and is reflected back 
when it reaches the other side. 
(Therefore this method is also called 
the reflection method or wall method.) 
The simulation domain is taken in the downstream rest frame, 
and a shock wave is excited due to the interaction between 
the reflected and the injected plasma. 
Additional methods, such as 
the flow-flow method (Omidi and Winske, 1992) 
and the plasma release method (Ohsawa, 1985), are
used in hybrid and full particle simulations.
A detailed description of each method is given in 
Lembege (2003). 

In these methods, 
an excited shock wave propagates upstream. 
Therefore,
it is necessary to take a very long simulation domain 
in the propagation direction of the shock wave 
in order to study a long-time evolution of the excited shock wave. 
This makes it difficult to perform multidimensional simulations 
even with a present-day supercomputers. 
In this letter, we first attempt to 
perform a full particle simulation of 
a collisionless shock in the shock rest frame: 
the collisionless shock is excited 
by using the ``relaxation method'' 
which was used in hybrid code simulations in 1980s 
(Leroy \textit{et al.}, 1981, 1982; Kan and Swift, 1983). 
This method has not been used in full particle simulations 
due to several difficulties in numerical techniques.

\section{Simulation Model}

% Code

Our simulation code (Umeda, 2004) is 
an improved version of one-dimensional 
Kyoto university electromagnetic particle code 
(Omura and Matsumoto, 1993), 
where Maxwell's equations and the relativistic equations of motion 
for individual electrons and ions 
are solved in a self-consistent manner. 
The continuity equation for charge is also 
solved to compute the exact current density given by 
the motion of the charged particles (Umeda \textit{et al.}, 2003).

% Model

The simulation domain is taken 
in a one-dimensional system along the $x$-axis. 
The initial state consists of two uniform regions 
separated by a discontinuity. 
In the upstream region that is taken in the left hand side 
of the simulation domain, 
electrons and ions are distributed uniformly in $x$ and 
are given random velocities $(v_x,v_y,v_z)$ to approximate 
shifted Maxwellian momentum distributions 
with the drift velocity $u_{x1}$, 
number density $n_{1} \equiv \epsilon_0 m_e \omega_{pe1}^2 / e^2$, 
isotropic temperatures $T_{e1} \equiv m_e v_{te1}^2$ and 
$T_{i1} \equiv m_i v_{ti1}^2$, 
where $m$, $e$, $\omega_{p1}$, and $v_{t1}$ are 
the mass, charge, upstream plasma frequency, and 
upstream thermal velocity, respectively. 
In this letter, subscripts ``1'' and ``2'' denote
``upstream'' and ``downstream'', respectively.
The upstream magnetic field $B_{0y1} \equiv -m_e \omega_{ce1}/e$ 
is also assumed to be uniform, where $\omega_{c1}$ 
is the upstream cyclotron frequency (with sign included). 
The downstream region taken in the right-hand side 
of the simulation domain is prepared similarly with 
the drift velocity $u_{x2}$, density $n_{2}$, 
isotropic temperatures $T_{e2}$ and $T_{i2}$, 
and magnetic field $B_{0y2}$. 
In this letter we assume a perpendicular shock (i.e., $B_{0x}=0$). 
As a motional electric field, an external electric field 
$E_{0z} =-u_{x1}B_{0y1} =-u_{x2}B_{0y2}$ 
is also applied in both the upstream and downstream regions. 
At the left boundary of the simulation domain,
we inject plasmas with the same quantities 
as those in the upstream region, 
while plasmas with the same quantities as those 
in the downstream region are also injected from the right boundary.
We adopted absorbing boundaries 
to suppress the non-physical reflection of electromagnetic waves at 
both ends of the simulation domain (Umeda \textit{et al.}, 2001). 

% Parameters

In the present simulation, 
the time, velocity, and position are 
normalized by the initial upstream electron plasma frequency 
$\omega_{pe1}=1.0$, 
upstream electron thermal velocity $v_{te1}=1.0$, and 
upstream electron Debye length 
$\lambda_{e1} \equiv v_{te1}/\omega_{pe1} = 1.0$, 
respectively. 
The initial temperatures in both upstream and downstream regions 
are assumed to be isotropic. 
In the upstream region,
we assume a low beta and weakly magnetized plasma, 
such that $\beta_{e1}=\beta_{i1}=0.125$, and 
$\omega_{ce1}/\omega_{pe1}=-0.05$, which are 
similar to the recent full particle simulations 
(Shimada and Hoshino, 2000;  Schmitz \textit{et al.}, 2002a, b; 
 Lee \textit{et al.}, 2004). 
The light speed in the present simulation 
is given as $c=80.0$. 
The bulk flow velocity of the upstream plasma is assumed to be 
$u_{x1}=4.0$, which corresponds to 
the Alfv\'{e}n Mach number $M_A = 10.0$. 
The ion-to-electron temperature ratio,
$r_T \equiv T_i/T_e$ is assumed to be unity 
in the upstream region ($r_{T1}=1$).
In addition to the upstream quantities $u_{x1}$, $\omega_{pe1}$, 
$\omega_{ce1}$, $v_{te1}$, and $r_{T1}$, we need the downstream 
ion-to-electron temperature ratio, $r_{T2}$,
so as to uniquely determine the other downstream quantities 
$u_{x2}$, $\omega_{pe2}$, $\omega_{ce2}$, and $v_{te2}$ 
from the shock jump conditions for a 
magnetized two-fluid plasma consisting of 
electrons and ions with the equal bulk velocity and 
the equal number density. 
We adopt $r_{T2}=4.0$ so that
the thermal velocities of both downstream electrons and ions 
become much slower than the light speed. 
However, note that we can choose an arbitrary value for $r_{T2}$.
As can be seen later, the final cyclic reformation state 
does not depend on the choice of $r_{T2}$.
Since we performed the present simulation on a personal computer,
we used a reduced ion-to-electron mass ratio $r_m =100$ 
for computational efficiency.
With these parameters, we obtain the initial downstream quantities 
as $\omega_{pe2}=1.95$, $\omega_{ce2}=-0.19$, 
$u_{x2}=1.05$, and $v_{te2}=7.55$.

\begin{figure}[t]
\includegraphics[width=20pc,height=40pc]{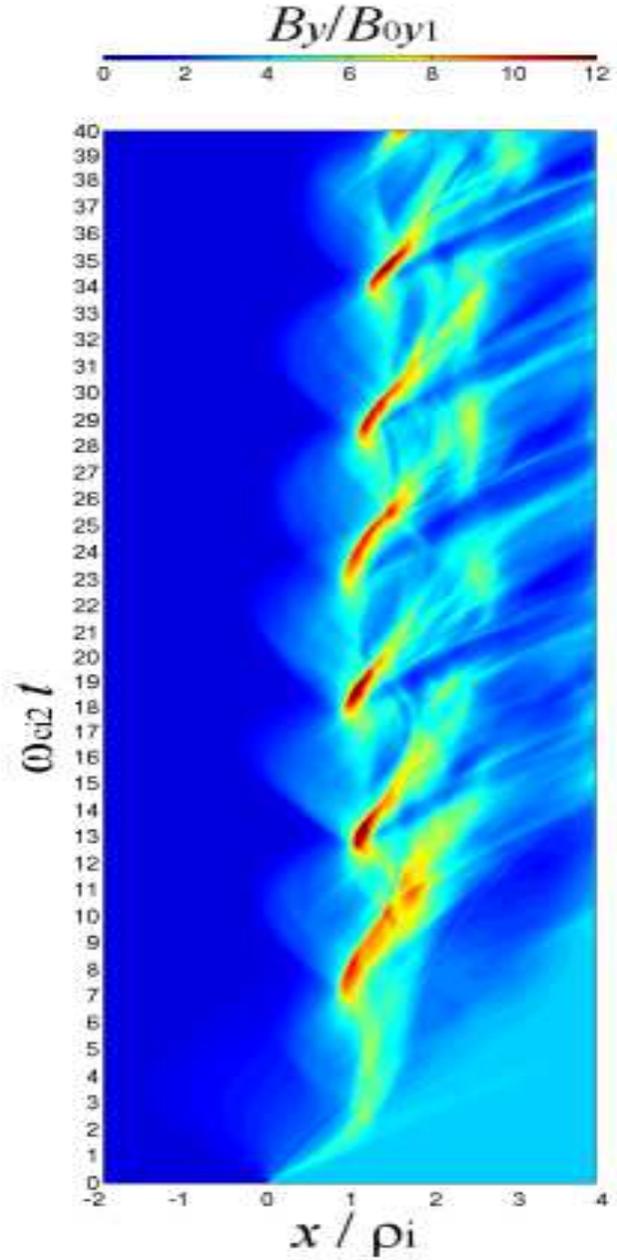}
\caption{The transverse magnetic field $B_y$ 
as a function of position and time. 
The time and position are normalized by 
$\omega_{ci2}$ and 
$\rho_i \equiv u_{x1}/\omega_{ci2}$, respectively. 
The magnitude of magnetic field is normalized by $B_{0y1}$. 
}
\end{figure}

We used 4096 cells for the upstream region and 
8192 cells for the downstream region.
The grid spacing and time step of the present simulation are 
$\Delta x/\lambda_{e1} = 1.0$ and $\omega_{pe1}\Delta t=0.01$.
We used 128 pairs of electrons and ions per cell in the upstream region 
and 512 pairs of electrons and ions per cell in the downstream region.
It should be noted that such a few number of particles per cell 
is not enough to suppress the enhanced thermal fluctuations of 
particle-in-cell codes. 
In the present simulation, however, 
the numerical noises due to random motions of individual particles 
are substantially reduced by 
adopting second-order schemes (Umeda, 2004).

\section{Simulation Result}

%%% Results 

\begin{figure*}[t]
\includegraphics[width=40pc,height=20pc]{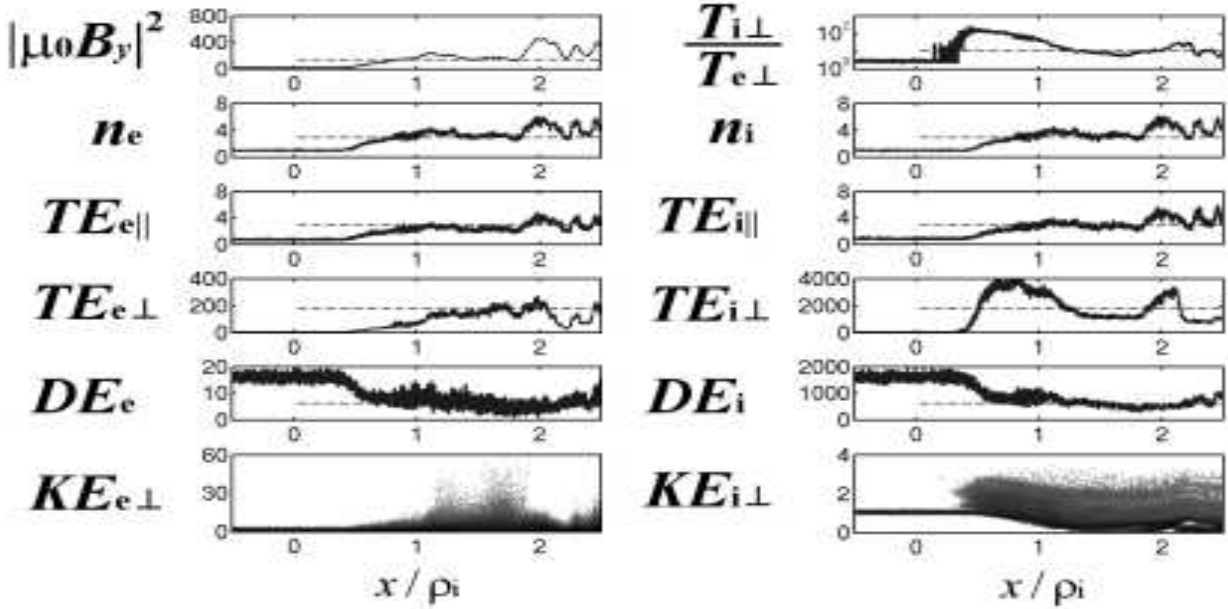}
\caption{
Spatial profiles of the magnetic field energy density 
$|\mu_0 B_y|^2$, 
ion-to-electron temperature ratio $r_T$, 
number densities $n$, thermal energy density components 
parallel and perpendicular to the ambient magnetic field 
$TE_{||}$ and $TE_{\perp}$, 
the drift energy density $DE$, and 
perpendicular kinetic energy versus position phase-space diagrams 
for electrons and ions at $\omega_{ci2}t = 38.1$. 
The number densities are normalized by the initial upstream 
electron density $n_{e1}$, 
and the energy densities are normalized by the initial upstream
thermal energy density of electrons $n_{e1}m_ev_{te1}^2$. 
The dashed lines show the downstream values obtained from 
the shock jump conditions for anisotropic plasmas 
with the downstream quantities 
$v_{te||2}=v_{te1}$, $r_{T||2}=r_{T1}$, $r_{T\perp2}=10$, 
and the shock speed $U/u_{x1}=0.0282$. 
The perpendicular kinetic energies for electrons and ions are 
normalized by their upstream bulk energies, i.e., 
$\frac{1}{2}m_e u_{x1}^2$ and $\frac{1}{2}m_i u_{x1}^2$ 
respectively. 
}
\end{figure*}

%   Magnetic field

Figure 1 shows the transverse magnetic field $B_y$ 
as a function of position and time.
The discontinuity exists at $x=0$ initially. 
However, the shock front shifts downstream 
due to the cyclotron motion of upstream ions 
that penetrate into the downstream region. 
A new shock front appears at $x/\rho_i \simeq 1$. 
The period from $\omega_{ci2}t =$ 0 to 7 
is the transition epoch from the initial to the 
 cyclic reformation state (see below). 
It should also be noted that 
the initial discontinuity causes the non-physical generation 
of electromagnetic waves at the onset 
because Ampere's law $\frac{\partial B_y}{\partial x}=\mu_0 J_z$ 
is not satisfied at the initial state. 
However, the electromagnetic waves are perfectly absorbed 
at both boundaries, 
and they  never affect the simulation result in later time.

The typical processes for the self-reformation of 
perpendicular collisionless shocks are as follows
(Quest, 1985; Lembege and Savoini, 1992; 
 Schmitz \textit{et al.}, 2002a, b; Nishimura \textit{et al.}, 2003; 
 Hada \textit{et al.}, 2003). 
A portion of ions are reflected back from the shock ramp. 
The shock foot region broadens toward the upstream region
as the reflected ions penetrate. 
However, the reflected ions return to the downstream region 
because of the cyclotron motion. 
The reflected ions can interact with upstream ions 
resulting in the self-formation of a new shock ramp 
in the old shock foot region. 
Finally, a new shock front appears at $x/\rho_i \simeq 1$, 
and ions are again reflected toward the upstream region 
from the new shock ramp. 

The present simulation result shows that the timescale of the 
self reformation is almost equal to 
the cyclotron period of the downstream ions 
($\omega_{ci2}t \simeq 2\pi$), which is
in agreement with the previous studies
%% showing that the shock self-reforms on a timescale of 
%% $\omega_{ci1}t = 1 \sim 2$ 
(Quest, 1985, Lembege and Savoini, 1992; 
 Schmitz \textit{et al.}, 2002a, b; Nishimura \textit{et al.}, 2003; 
 Scholer \textit{et al.}, 2003; 
 Lee \textit{et al.}, 2004).
Scholer \textit{et al.} (2003) showed that 
the self reformation process is not a computational artifact 
with the real ion-to-electron mass ratio, 
while the ratio of electron plasma-to-cyclotron frequency 
is smaller ($\omega_{pe1}/\omega_{ce1}<10$). 
On the other hand, 
Lee \textit{et al.} (2004) generated 
the self reformation of more than six cycles, 
while their mass ratio is much smaller ($r_m=20$). 
In the present shock rest frame, 
we have also confirmed the shock reformation process 
up to six cycles with $r_m=100$, but with less grid cells.

% Downstream Conditions

In order to analyze the downstream condition 
in the cyclic reformation state, we plot in Figure 2 
spatial profiles of the magnetic energy density $|\mu_0 B_y|^2$, 
the ion-to-electron temperature ratio 
perpendicular to the magnetic field $r_{T\perp}$, 
the number density $n$,  parallel and perpendicular 
thermal energy density components $TE_{||}$ and $TE_{\perp}$, 
and the drift energy density $DE$ 
for electron and ions at $\omega_{ci2}t=38.1$.

In the present shock-rest-frame model, 
a shock transition layer is self-consistently formed 
due to relaxation of the two plasmas with different quantities. 
The downstream region at the resulting cyclic reformation state 
is quite different from that of the initial state. 
For both electrons and ions, 
the spatial profiles of the thermal energy density 
component parallel to the ambient magnetic field 
are similar to those of number densities. 
This means that 
the downstream parallel temperatures for both electrons and ions 
become almost the same as those in the upstream region, 
i.e., $T_{e||1}\simeq T_{e||2}\simeq T_{i||1}\simeq T_{i||2}$. 
On the other hand, 
the ratio of the ion-to-electron perpendicular temperature ratio 
is very large ($\sim10^2$) in the transition region, and 
is typically $r_{T\perp2} \sim 10$ in the downstream region 
($r_{T\perp2} = 5.0 \sim 20.0$).

As seen in Figure~1,
the excited shock wave propagates slowly downstream 
with the roughly estimated shock speed $U/u_{x1}=0.0282$. 
The downstream electron and ion number densities 
in the cyclic reformation state become 
smaller than those at the initial state, whereas 
the downstream electron and ion bulk velocities 
become faster than those at the initial state. 
We performed several additional runs with different system sizes
and found that the simulation results --- i.e., 
the shock speeds and 
the spatial profile of all physical quantities at 
an arbitrary time --- are almost unchanged. 
In other words, the present system size 
is long enough to discuss the kinetic processes 
in the shock transition region.

In the downstream region of the cyclic reformation state,
the physical quantities fluctuate and are not spatially uniform. 
We solved the shock jump conditions for anisotropic plasmas 
(Hudson, 1970) as a reference. 
Taking into account the shock speed 
and the typical downstream quantities 
$v_{te||2}=v_{te1}$, $r_{T||2}=r_{T1}$, and $r_{T\perp2}=10$, 
the other downstream quantities are obtained as 
$\omega_{pe2}=1.71$, $\omega_{ce2}=-0.146$, 
$u_{x2}=1.33$, and $v_{te\perp2}=5.42$. 
These quantities are plotted in Figure 2 with dashed lines. 
A difference between the quasi-steady state 
by the shock jump conditions and the simulated downstream state 
is because of the dynamical shock reformation process.  
This might be another reason why we did not obtain 
the rigorous shock rest frame.

% Discussion

For a given set of upstream parameters, 
the fluid shock jump conditions cannot give 
the downstream state uniquely. 
In the electron-ion fluid, 
the total plasma pressure is defined as 
the sum of the electron and ion pressures, $P=n ( T_e+ T_i)$, 
and the shock jump conditions allow us to 
take an arbitrary downstream temperature ratio, $r_{T2}$. 
However, 
the value of $r_{T2}$ is determined by 
the kinetic dynamics as seen in the simulation result.
We performed several additional runs with different
initial downstream values and confirmed that
the shock speed in our reference frame
and downstream thermal properties at
the cyclic reformation state 
do not depend on the initial value of $r_{T2}$
but on the other initial downstream parameters, such as 
the magnetic field, the number density, the bulk velocity, 
and the total plasma pressure.

In the bottom panels of Figure 2 we show 
perpendicular kinetic energy versus position 
diagrams for electrons and ions. 
We found that 
there exists a supra-thermal component of electrons 
at two local areas. 
The previous works reported 
the electron surfing acceleration with electrostatic solitary waves 
in the transition region 
(Hoshino and Shimada, 2002; 
 Schmitz \textit{et al.}, 2002a, b). 
In the present simulation, we confirmed 
the existence of solitary waves 
in the shock foot region, where 
the maximum kinetic energy is 
about 30-fold more than  the initial kinetic energy. 
On the other hand, 
another supra-thermal component of electrons 
due to the ion deceleration is seen at the overshoot, 
where the maximum kinetic energy is much more ($ > 60 KE_{e1}$) 
than that of 
the non-thermal electrons via the surfing mechanism. 
Although this process was also found in the previous simulations 
(Hoshino and Shimada, 2002; 
 Schmitz \textit{et al.}, 2002a, b), 
the maximum kinetic energy in the present simulation  
is more than that in their simulations. 
We expect that 
this process becomes more significant with a larger mass ratio.

\section{Conclusion}

We have developed a shock-rest-frame model 
for full particle simulations of perpendicular collisionless shocks 
based on the relaxation method 
used in the previous hybrid simulations 
(e.g., Leroy \textit{et al.}, 1981, 1982). 
We reconfirmed 
both the formation of microscopic solitary structures 
due to the current-driven instability 
and cyclic reformation for a long time 
with a much smaller simulation domain. 
%We also found a strong electron energization process 
%at the shock overshoot. 
The shock-rest-frame model allows us to perform 
multidimensional full particle simulations of planar shocks 
more readily with current supercomputers. 
We are extending 
the present shock-rest-frame model to oblique shocks. 
Preliminary results show that the present rest-frame model 
is very useful for exciting shock waves with arbitrary parameters.

\acknowledgments{
The authors are grateful to F.~Takahara, N.~Okabe, and T.~N.~Kato 
for discussions. 
The computer simulation was performed as a collaborative research project 
at STEL in Nagoya University and at YITP in Kyoto University.
This work was supported by Grant-in-aid 
for Encouragement of Young Scientists (B) \#18740153 
from the Japan Ministry of Education, Culture, Sports, Science, 
and Technology (R.~Y.). 
}

%%%%%%%%%%%%%%%%%%%%%%%%%%%%%%%%%%%%%%%%%%%%%%%%%%%%%%%%%%%%

\email{
T. Umeda (email: umeda@stelab.nagoya-u.ac.jp) and  
R. Yamazaki (email: ryo@theo.phys.sci.hiroshima-u.ac.jp)}

\label{finalpage}
\lastpagesettings

\end{document}